**Research Article**

Javier Rodríguez-Álvarez*, Joan Vila-Comamala, Antonio García-Martín, Albert Guerrero, Xavier Borrisé, Francesc Pérez-Murano, Christian David, Alvaro Blanco, Carlos Pecharromán, Xavier Batlle, Arantxa Fraile Rodríguez, and Amílcar Labarta.

# Dichroism of coupled multipolar plasmonic modes in twisted triskelion stacks

**Abstract:** We present a systematic investigation of the optical response to circularly polarized illumination in twisted stacked plasmonic nanostructures. The system consists in two identical, parallel gold triskelia, centrally aligned and rotated at a certain angle relative to each other. Sample fabrication was accomplished through a double electron beam lithography process. This stack holds two plasmonic modes of multipolar character in the near-infrared range, showing a strong dependence of their excitation intensities on the handedness of the circularly polarized in incident light. This translates in a large circular dichroism which can be modulated by adjusting the twist angle of the stack. Fourier-transform infrared (FTIR) spectroscopy and numerical simulations were employed to characterize the spectral features of the modes. Remarkably, in contrast to previous results in other stacked nanostructures, the system's response exhibits a behavior analogous to that of two interacting dipoles only at small angles. As the angle approaches 15°, where maximum dichroism is observed, more complex modes of the stack emerge. These modes evolve towards two in-phase multipolar excitations of the two triskelia as the angle increases up to 60°. Finally, simulations for a triangular array of such stacked elements show a sharp mode arising from the hybridization of a surface lattice resonance with the low-energy mode of the stack. This hybridized mode demonstrates the capability to be selectively switched on and off through the light polarization handedness.

\* **Corresponding author: Javier Rodríguez-Álvarez,** Departament de Física de la Matèria Condensada, Universitat de Barcelona, 08028 Barcelona, Spain and Institut de Nanociència i Nanotecnologia (IN2UB), Universitat de Barcelona, 08028, Spain.
E-mail: javier.rodriguez@ub.edu
**Xavier Batlle, Arantxa Fraile Rodríguez, and Amílcar Labarta,** Departament de Física de la Matèria Condensada, Universitat de Barcelona, 08028 Barcelona, Spain and Institut de Nanociència i Nanotecnologia (IN2UB), Universitat de Barcelona, 08028, Spain.
**Joan Vila-Comamala and Christian David,** Paul Scherrer Institute, Forschungsstrasse 111, Villigen 5232, Switzerland.
**Antonio García-Martín,** Instituto de Micro y Nanotecnología IMN-CNM, CSIC, CEI UAM + CSIC, Isaac Newton 8, 28760, Tres Cantos, Madrid, Spain.
**Albert Guerreo and Francesc Pérez-Murano,** Institut de Microelectrónica de Barcelona (IMB-CNM, CSIC), Bellaterra, 08193, Spain.
**Xavier Borrisé,** Catalan Institute of Nanoscience and Nanotechnology (ICN2), CSIC and BIST, Campus UAB, Bellaterra, 08193 Barcelona, Spain.
**Alvaro Blanco and Carlos Pecharromán,** Instituto de Ciencia de Materiales de Madrid (ICMM), Consejo Superior de Investigaciones Científicas (CSIC), Calle Sor Juana Inés de la Cruz 3, Madrid, E-28049 Spain.

## 1 Introduction

Chirality is an intrinsic geometrical property of objects that cannot be superimposed with their mirror image. This property gives rise to two distinct realizations, termed enantiomers, which exhibit opposite handedness. Although often overlooked, chirality plays a crucial role in numerous critical phenomena, particularly in the field of biochemistry. The handedness of biological molecules such as amino acids, DNA or enzymes, for instance, has been demonstrated to be of paramount importance in elucidating their properties [1], [2]. A remarkable example is limonene, where one enantiomer is responsible for the characteristic fragrance of citrus fruits, while its counterpart contributes significantly to the scent of numerous coniferous and broadleaved trees [3].

Chirality is not limited to physical objects but can also be a property of fields, as is the case with left-handed (LCP) and right-handed (RCP) circularly polarized light [4], [5]. Light with a specific handedness serves as a valuable probe for chiral media, since light-matter interaction can be dictated by the handedness of both entities [6], [7], when present. Such chiroptical activity is commonly quantified through the optical functions $f$ of the system under the two circular polarizations by means of the circular dichroism (CD) that can be defined as:

$$CD_f = \frac{f_{LCP} - f_{RCP}}{f_{LCP} - f_{RCP}}, \quad (1)$$

where $f$ stands for any of the optical cross-sections (CS), namely, the absorption, scattering, and extinction, under either RCP or LCP illumination.

The chiral interaction in naturally occurring systems is typically of low magnitude, presenting substantial challenges for detection and quantification. Consequently, engineered chiral plasmonic structures have gained relevance because of their enhanced light-matter interaction associated with large values of the near fields, showing enhancements of the chiral signal of several orders of magnitude [8]–[11]. Furthermore, their integration into hybrid nanostructures promises active control of chirality [12].

Nowadays, strategies to fabricate chiral nanostructures are very diverse. However, due to the inherent difficulty to manufacture 3D structures using conventional nanofabrication techniques, early approaches were based on planar plasmonic structures [13]–[17]. These planar configurations typically exhibited limited chiroptical activity due to their two-dimensional nature, as truly planar structures cannot manifest chiroptical activity without violating system reciprocity [18]. Consequently, the observed chiroptical activity in many such structures was primarily attributed to their interaction with the substrate, effectively breaking the planar approximation [14], [16], [19]. Alternative designs have relied on the interaction between two modes through the excitation of Fano resonances [20], [21]. While these systems present circular dichroism (CD) in the absorption and the scattering cross sections (CS), the opposite sign of those contributions result in mutual cancellation. As a result, the extinction CS remains polarization-independent, again in accordance with reciprocity principles [22].

Further developments in nanofabrication technology have enabled the realization of diverse 3D nanostructures showing large values of CD, such as spirals [23], helicoidal arrangements [24], [25], and various stacked planar nanostructures ranging from nanorods [26], crosses [27], split ring resonators [28], [29], and gammadions [30], [31], among others.

In this work, we present a twisted stack formed by two identical planar nanostructures, called triskelia, which are characterized by their inherent 2D chirality and threefold rotational symmetry. In contrast to previously reported systems [24]–[29], the monomers of the stack exhibit intrinsic chiroptical activity. Furthermore, their threefold symmetry hinders the excitation of simple modes with an even number of poles while promoting the mixing of higher-order multipolar modes. As elucidated in our previous work [32], the small distance between the stack elements facilitates strong inter-element interactions, resulting in the emergence of handedness-dependent resonances in the near infrared spectral region. This configuration gives rise to pronounced CD which can be modulated by adjusting the twist angle between the stacked elements. Here, we investigate the nature of those resonances as a function of the twist angle between the two stacked structures, demonstrating that higher-order multipolar modes play a crucial role in the excitation of the structure under circularly polarized illumination. Our study is based on Fourier transform infrared spectroscopy (FTIR) measurements and finite difference time domain (FDTD) simulations.

## 2 Triskelion stack

The motif constituting the building block of the stack, the triskelion, is designed as three identical gold elements extending from a common central point, with each element oriented at 120° intervals. These elements exhibit a clockwise bend of 120° at their respective midpoints, thereby preserving threefold rotational symmetry while eliminating mirror planes parallel to the axis of rotation (see Fig. 1a). This geometric arrangement confers 2D chirality to the triskelion structure. The specific dimensions and geometry of the studied monomer are shown in Fig. 1a. The thickness of the planar structure was set to 30 nm. The final 3D structure consisted of a twisted stack of two coupled plasmonic triskelia, a schematic of which is shown in Fig. 1b, where the twist angle $\alpha$ is defined so that the stack has right-handed chirality whenever $\alpha < 60°$. Note that the chirality of the stack is the opposite (left-handed) for $\alpha > 60°$ due to the threefold symmetry of the monomer. $\alpha = 60°$ is a symmetrical configuration without a distinct handedness.

Samples with various twist angles were manufactured through sequential processes by electron beam lithography (EBL), all of them with a separation between the two stacked elements of about 20 nm. Additional information regarding the manufacturing process is shown in the Methods section. Some examples of the manufactured stacks are shown in Fig. 1c. The cross-sectional image in Fig. 1d indicates good parallelism and consistent separation between the two triskelia. To preserve optical equivalence across all interfaces of the structure, the stack was embedded in a homogeneous medium with a refractive index of $n=1.42$. Larger areas of some samples are shown in Fig. 1e and f, illustrating the uniformity of the stacks.

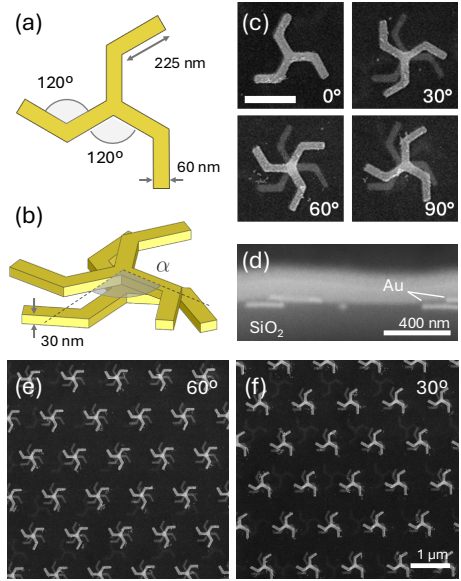

**Fig. 1:** (a) Diagram of a triskelion displaying the dimensions of the studied case. (b) Schematic depiction of the stack indicating the twist angle $\alpha$, the vertical separation in the manufactured samples is 20 nm approximately. (c) Scanning Electron Microscopy (SEM) top-view images of some of the fabricated nanostructures. (d) Cross-section SEM image of one of the samples. Scale bars are 400 nm in length. (e) and (f) correspond to SEM images showing the spatial arrangement of the fabricated array of stacked triskelia. Note that the fabricated array has a pitch of 1200 nm.

## 3 Results and discussion

The structure exhibits two plasmonic resonances within the wavelength range of 1100 and 1600 nm. Both, FTIR measurements and simulations reveal two extinction peaks in this range, which undergo remarkable spectral shifts and intensity as the twist angle is modified. This is illustrated in Fig. 2a and b, where the intensity of the peaks, particularly the one at lower energy, is severely affected by the handedness of the circular polarization of the incident radiation. As a result, these handedness-dependent resonances yield large CD, as shown in Fig. 2c and d. It is worth noting that, while the studied excitations are of multipolar character and not as intense as the primary dipolar excitations (far in the infrared range, around 2000-3000 nm, for this structure), there is a general qualitative agreement between the experimental and simulated curves in Fig. 2, including the opposite sign of the CD observed for the 30° and 90° cases.

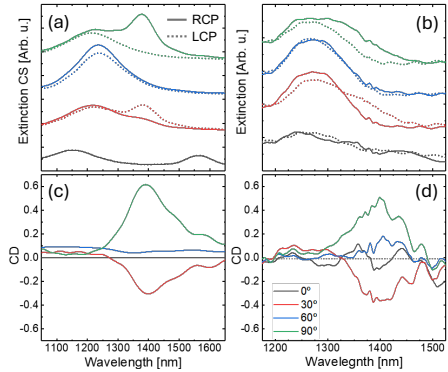

**Fig. 2:** (a) Simulated and (b) measured extinction spectra for different twist angles under LCP (solid line) and RCP (dashed line) illumination. The baselines of the spectra are shifted arbitrarily for clarity. (c) Simulated and (d) experimental CD extracted from the data in panels (a) and (b), respectively, using Eq. (1).

Moreover, owing to the threefold symmetry of the triskelion, both the spectral shifts and intensity variations are almost symmetric around 60° when the handedness of the light polarization is exchanged. Thus, the spectra for 30° and 90° under LCP and RCP light, respectively, appear remarkably similar (see Fig. 2a and b). The differences in the relative intensity of the peaks at these two angles can be attributed to the rightward bending of the triskelion branches, which breaks the symmetry when switching the light polarization between the 30° and 90° configurations. This also accounts for the minor differences observed in the spectra under LCP and RCP light at 60°, resulting in CD that is not strictly null, contrary to what would be expected solely from the threefold symmetry of the triskelion.

The typical approach to describing the optical response of a chiral medium would be to consider a simple and straightforward framework: the Born-Kuhn model [33]–[35]. This approximation supposes two interacting electrons oscillating in orthogonal directions in parallel planes separated a distance $d$. The coupling between the two particles results in two new eigenstates of the coupled oscillators system. This model yields results analogous to those of the hybridization of two plasmonic modes [36], [37] and was experimentally shown to be valid for plasmonic systems [38]. However, in prior studies, $d$ was of the order of the excitation wavelength, while in the present work, $d \sim 20$ nm. As a result, the phase shift of the incident radiation between the two monomers is rather small, weakening the chiral response. At the same time, since the interaction between plasmonic elements is mediated by their near fields, the close proximity between the two monomers in our study significantly enhances their coupling [26]. Consequently, a model based on the phase shift alone, such as the Born-Kuhn model, cannot account for the optical response of the triskelion stack. We propose instead an interpretation based on the hybridization of the plasmonic modes of individual triskelia. As shown in the charge distributions snapshots of the triskelia in Fig. 3, the complexity of the motif gives rise to multipolar excitations, especially fostered by geometric frustration; specifically, the threefold symmetry of the triskelion does not effectively accommodate an even number of poles, which are typically associated with simple dipole or multipolar modes. These plasmonic resonances, in turn, interact with those of the second triskelion in the stack resulting in a plethora of hybrid modes that deviate from the ideal in-phase/anti-phase excitations often considered in previous works involving two interacting dipoles [16], [29]. Moreover, the geometry of the system changes with the twist angle, favoring the excitation of low-energy resonances displaying variable dephasing between the polarizations of the two triskelia. This is clear in Fig. 3c and Fig. S1 in the Supplementary Information, where charge distributions and computed dipole moments for the two components of the stack as a function of α for the high- and low-energy modes are shown. Thus, while the high-energy mode maintains the two electric dipole moments relatively in-phase as α changes, the low energy mode exhibits a variable dephasing between the polarizations of both triskelia in the stack. Only at very small angles, the low-energy mode is sufficiently close to the anti-phase polar configuration (see Fig. 3a, Fig. 3c and Fig. S1 in the Supplementary Information). Around 30°, the polarizations of the two triskelia become perpendicular in the low-energy mode (see Fig. 3b, Fig. 3c and Fig. S1 in the Supplementary Information). As the angle α further increases, the polarizations corresponding to the low- and high-energy modes begin to resemble each other more closely (see Fig. 3c and Fig. S1 in the Supplementary Information). This observation indicates that, although only two peaks are detected for any value of α, the features of the low-energy mode excited in each case are significantly different.

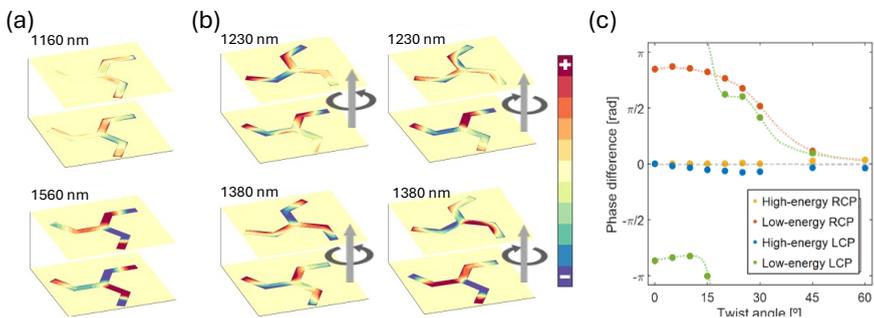

**Fig. 3:** Simulated charge distributions for facing triskelion interfaces for twist angles 0° (a) and 30° (b). Simulations at different angles correspond to arbitrary values of the phase of the incoming illumination. Panel (c) presents the phase difference in the orientations of the electric dipole moment of the top and bottom triskelia for each twist angle. All the information displayed shares a common scale. Note that a single polarization is presented for the case of 0° since the response is identical for both LCP and RCP. The geometry of the illumination is indicated by the vertical gray arrows and corresponds to that of the FTIR measurements.

These findings suggest that the chiroptical activity in the stack arises from the dependence of the excitation intensity of these modes on the light-handedness, in accordance with previously reported results [26], [27], [32]. However, in this case, the resonances yielding CD are not sufficiently explained by the simplistic model involving the splitting of a triskelion mode into anti-phase and in-phase modes of the stack. Instead, they are better characterized by the excitation of an out-of-phase mode that is highly sensitive to the geometry of the stack at each angle α, in conjunction with the in-phase excitation. This continuous shift of the dephasing between the dipole moments of the monomers in the stack is shown in Fig. 3 and Fig. S1 in the Supplementary Information, for angles between 0° and 60°. In addition, the small net dipole moment of the out-of-phase, low-energy resonances agree with the fact that CD is typically associated with the selective excitation of modes exhibiting poor scattering [32].

In Fig. 4, the spectral response of the stack as a function of α is shown in colormaps obtained from FTIR and FDTD simulated data. Remarkably qualitative agreement between both sets of colormaps is found. Two Lorentzian functions, corresponding to the low- and high- energy resonances, were fitted to both the experimental and simulated spectra, with the position of some of these peaks also displayed for comparison in Fig. 4. The details of the fitting are presented in Fig. S2 in the Supplementary Information.

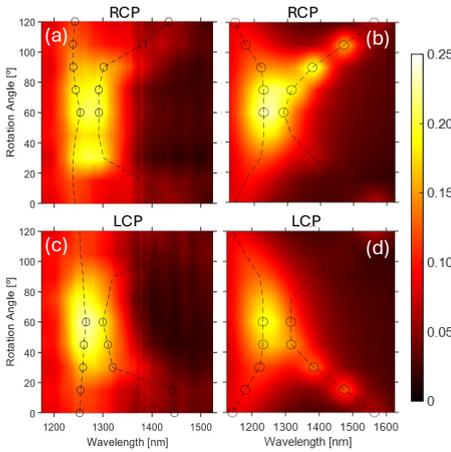

**Fig. 4:** Colormaps depicting the experimental ((a), (c)) and simulated ((b), (d)) extinction spectra for twist angles between 0° and 120° for both RCP ((a), (b)) and LCP ((c), (d)). Empty circles highlight the spectral position of the two Lorentzian peaks fitted to both the experimental and simulated data. Dashed lines are guidelines to the eye.

For angles between 0° and 60°, the structure favors the excitation of the out-of-phase, low-energy mode for the light-handedness opposite to that of the stack, i.e. LCP since the chirality of the stack is right-handed. Consequently, the out-of-phase mode is most efficiently excited by circular polarization that induces the greatest change in the relative orientation of the electric field with respect to the lower and upper triskelia (note that illumination is from bottom to top). In contrast, RCP light will excite both triskelia similarly due to the minimal phase shift of the light across $d$, thereby favoring the in-phase mode and significantly hindering the out-of-phase resonance. At 60°, the excitation features of the system are quite similar under both circular polarizations of the light, with only slight variations arising from the rightward bending of the triskelion branches. Despite this minor splitting at this angle, the intensity of the low-energy resonance is nearly negligible, as shown in Fig. S2 of the Supplementary Information, resulting in a single prominent peak in the extinction spectra. For angles greater than 60°, symmetric optical responses are observed with respect to those for α < 60° under reversed circular polarizations, except for small differences in the intensity of the low-energy mode. This slight asymmetry is owed to the fact that the triskelion is a chiral object in 2D, thus slightly altering the expected symmetric response around 60° due to its inherent threefold symmetry.

Finally, it is important to highlight that arranging these twisted triskelion stacks in a triangular array with an appropriate pitch (see Fig. 1e and f for details) can enhance the excitation of a surface lattice resonance (SLR). The intensity of this resonance strongly depends on the handedness of the circular polarization of the light and the twist angle, resulting in very large dichroism for angles larger than 10°. These results were obtained from numerical simulation of a lattice embedded in a medium of $n = 1.5$, where each element consisted of a twisted stack with 50 nm in separation between the two monomers and a twist angle range of $0° ≤ α ≤ 30°$. Fig. 5 shows the simulated extinction CS under RCP an LCP light at four selected values of α, with a fixed pitch of 1200 nm. At this pitch, a SLR emerges as a sharp peak with a Fano profile located around 1550 nm, precisely in the region where anti-phase excitations of the triskelion stack become energetically favorable (as seen in the case for α = 0° in Fig. 5). Therefore, the Fano resonance originates from the hybridization of the SLR mode with the anti-phase plasmonic excitation of the stack, where the sharp and intense peak of the resonance corresponds to the coherent excitation of the anti-phase mode across the lattice of triskelion stacks through absorption of the incoming radiation. Consequently, the cooperative interaction among the near fields created by the elements in the lattice, due to the geometrical resonance condition, amplifies the excitation of the anti-phase mode in every stack (which would otherwise be very weak), leading to a large energy absorption. This excitation occurs with much greater efficiency when illuminated by light that matches the handedness of the stack (LCP in this case), as incoming radiation with the handedness of the stack favors a larger polarization of the first triskelion. This, in turn, induces a larger anti-phase polarization in the second triskelion through near-field interactions. It is important to note that in-phase modes cannot be excited in this range of relatively low energies. Under circular polarization with opposite handedness to that of the stack (RCP in this case), polarizations induced in both triskelia within each stack are reduced, leading to lower absorption associated with the lattice resonance. As a result, the spectra in Fig. 5 exhibit large differences in the intensity of the SLR modes under RCP and LCP illumination, resulting in high values of dichroism, except at $\alpha = 0°$ where the stack lacks chirality.

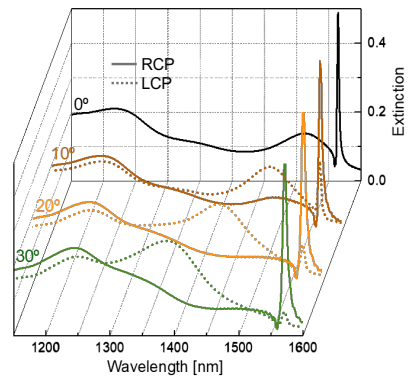

**Fig. 5:** Simulated extinction spectra for an array of stacked triskelia with a pitch of 1200 nm, embedded in a homogeneous medium of n=1.5 with a separation of 50 nm between triskelia. The response of the array under RCP and LCP illumination is presented for different values of the twist angle within the stack.

We attempted to experimentally reproduce these results using samples like those shown in Fig. 1e and f but were unsuccessful. This may be attributed to the narrow spectral width of the SLR, inadequate perpendicular illumination of the samples due to the numerical apertures of the optical systems used for light focusing, or insufficient patterned areas on the fabricated samples, which could lead to finite-size broadening effects.

## 4 Conclusions

We have presented a comprehensive study of the chiroptical response of a twisted stack of triskelion nanostructures, emphasizing the detailed characterization of two coupled multipolar modes in the near-infrared region. Our findings, supported by FTIR measurements and FDTD simulations, demonstrate that both the spectral position and intensity of the resonances can be precisely tuned by adjusting the angle between the two elements in the stack. The three-fold symmetry of the elements promotes complex multipolar excitations which are incompatible with an even number of pairs of equal poles with opposite sign. In particular, under illumination with an opposite handedness to that of the structure, two multipolar resonances are observed; conversely, illumination with light that matches the handedness of the structure reveals only the peak corresponding to the high-energy excitation. This results in large dichroism in the extinction CS of the system, except in the particularly symmetric cases at twist angles of 0° and 60°.

Our findings show that the high-energy mode originates from an in-phase excitation of the two triskelia under both RCP and LCP illumination. In contrast, the low-energy mode, which is enhanced by illumination under opposite handedness to that of the stack, corresponds to the excitation of the two monomers with a specific phase difference between their instantaneous polarizations. Thus, contrary to previous works, we demonstrate that anti-phase oscillations occur only at small twist angles; for angles greater than 15°, the polarizations of the two monomers progressively align in phase as α increases towards 60°. This draws a significantly more complex scenario depending on the twist angle, diverging from the descriptions provided by the Born-Kuhn model or the hybridization of two equivalent plasmonic resonances producing simple in phase- and antiphase-coupled modes of the stack.

Finally, simulations indicate that arranging such a chiral stack in a plasmonic triangular lattice leads to the selective excitation of a SLR depending on the light handedness. This configuration enables high CD values within a narrow wavelength range, that can be tuned by the pitch of the lattice and the geometry of the stack.

## 5 Methods

### 5.1 Sample fabrication

The samples were fabricated by consecutive EBL, following the steps depicted in Fig. S3 in the Supplementary Information. The samples were lithographed on a Si substrate (250 µm thick) coated with a $SiO_2$ layer on both sides (1.8 µm thick, deposited by wet thermal oxidation). The first EBL process aimed at defining the alignment markers, which consisted of Au squares of 100x100 nm$^2$ and 80 nm in thickness using an adhesion layer of 10 nm of Cr. The thickness of these markers has proven to be critical for their correct detection, and therefore, alignment of the EBL system for further lithography processes. The second step was to cover the markers with hydrogen silsesquioxane (HSQ) in solution in methyl isobutyl ketone (MIBK), commercially labelled as FOx (flowable oxide) [39], [40]. This diluted polymer was deposited through spin coating, pre-baked at 80 °C for 4 minutes and then annealed at 400 °C for a minimum of 1 hour. This produced a homogeneous layer of $SiO_2$ of 300 nm in thickness approximately following a cost-effective and controllable protocol.

Subsequently, an additional EBL step was carried out to pattern the first layer of triskelia, using the alignment markers as a reference. Following the corresponding exposure and development, 30 nm of Au was deposited on top of a 2 nm adhesion layer of Ge, which has been shown to better preserve the plasmonic properties of noble metal nanostructures compared to other common materials, such as Cr or Ti [41]. After completing the lift-off process, another layer of HSQ was deposited and then annealed, with this layer having a thickness of approximately 50 nm.

Finally, the third and final EBL step was performed. A second layer of twisted triskelia was patterned, ensuring that the centers of the triskelia in both layers were aligned, using the same alignment markers and parameters as in the previous step. Following this complex process, the two layers of twisted triskelia, separated by 20 nm, were embedded in $SiO_2$. Interestingly, despite the substrate consisting of a relatively thick layer of Si, the transparency of the entire double-oxidized Si wafer is approximately 50%, particularly for wavelengths larger than 1150 nm (see Fig. S4 in the Supplementary Information).

EBL was performed using a Raith/Vistec EBPG 5000Plus with a Gaussian-shaped beam, offering a maximum writing field size of 1024×1024 µm$^2$ and high beam stepping frequencies of up to 125 MHz, and operating at 100 keV acceleration and with an overlay precision around 20 nm for larger write fields and 10 nm for smaller ones (100×100 µm$^2$). The metallization was performed using physical vapor deposition (PVD) by evaporating the target using an electron gun.

### 5.2 FDTD simulations

The FDTD simulations presented in this work were performed using the commercial software provided by Lumerical [42]. All simulations were performed using a 2x2x2 nm$^3$ cell size setting a homogeneous background refractive index of 1.42 without absorption to better reproduce the optical properties of annealed HSQ. The optical coefficients for Au were obtained from [43]. The circularly polarized illumination was introduced using linearly polarized plane waves by setting two sources with orthogonal polarization angles and a π/2 phase difference between them. Two types of simulations were performed, those featuring single stacked elements, and others where the response of an infinite triangular lattice of stacked elements was simulated.

The former relies on the use of total-field scattered-field (TFSF) sources and perfect absorbing boundary conditions. Absorption and scattering CS data were collected since only one element was being simulated. For these simulations the extinction CS was the sum of absorption and scattering CS. The second type of simulations require the use of plane wave sources and periodic boundary conditions at the in-plane boundaries. In this case, reflection and transmission through the structure were measured. The extinction in this framework was assigned to be 1 - $T$, where $T$ stands for the transmission CS. It is worth noting that the reflection of the structure was very small compared to the extinction signal.

## 5.3 FTIR measurements

Bruker Vertex 70V Spectrophotometer with a microscope Hyperion 2000 was used to measure the extinction spectra. Transmission through the sample was measured using 4x magnification in order to minimize the numerical aperture of the system and to optimize normal incidence.

**Research funding:** Financial support from the Spanish Ministry of Science and Innovation (PID2021-127397NB-I00), the European Union FEDER funds, and the Generalitat de Catalunya (2021SGR00328) is acknowledged.

**Author contributions:** JRA has designed the structures, simulated the system, fabricated the samples, assisted in their characterization, and contributed to the draft and final manuscript. JVC and CD. have provided access to the facilities and the expertise necessary for the fabrication of the samples and have supported the process. AGM has contributed to the simulations and the discussion of the system. AGB, XB and FPM have contributed to the design and initial exploration of the fabrication process and the later characterization of the cross-section of the samples by SEM. AB and CP have performed the experimental characterization of the spectral response of the system by FTIR. XB and AFR have contributed to the design, discussion and revisions of the manuscript. AL has contributed to the design, discussion and elaboration of the draft and final manuscripts. All authors have accepted responsibility for the entire content of this manuscript and consented to its submission to the journal, reviewed all the results and approved the final version of the manuscript.

**Data availability:** The data that support the findings of this study are available from the corresponding author upon reasonable request.